\documentclass{aastex62}

\usepackage{graphicx}
\usepackage{graphics}
\usepackage{amssymb}
\usepackage{bm}
\usepackage{times}

\newcommand{\beq}{\begin{equation}}
\newcommand{\eeq}{\end{equation}}

\newcommand{\cm}{cm$^{-2}$}
\newcommand{\msun}{\textrm{M}_\odot}
\newcommand{\mmol}{\rm{M_{mol}}}
\newcommand{\mstar}{\rm{M_{\star}}}

\newcommand{\kms}{km~s$^{-1}$}

\newcommand{\hi}{H{\sc i}}
\newcommand{\nhi}{N_{\rm HI}}
\newcommand{\cii}{[C{\sc ii}]~158$\mu$m}
\newcommand{\hii}{H{\sc i}\,21cm}
\newcommand{\aco}{\alpha_{\rm CO}}

\shorttitle{Massive DLA galaxies at $z \approx 2$}
\shortauthors{Kanekar et al.}
\begin{document}
\title{High Molecular Gas Masses in Absorption-selected Galaxies at $z \approx 2$}

\correspondingauthor{Nissim Kanekar}
\email{nkanekar@ncra.tifr.res.in}

\author{N. Kanekar} 
\affiliation{National Centre for Radio Astrophysics, Tata Institute of Fundamental Research, 
Pune University, Pune 411007, India}

\author{J. X. Prochaska}
\affiliation{University of California Observatories-Lick Observatory, University of California, Santa Cruz, CA, 95064, USA}

\author{M. Neeleman}
\affiliation{Max-Planck-Institut f{\"u}r Astronomie, K{\"o}nigstuhl 17, D-69117 Heidelberg, Germany}

\author{L.~Christensen}
\affiliation{Dark Cosmology Centre, Niels Bohr Institute, Copenhagen University, Juliane Maries Vej 30, 2100 Copenhagen O, Denmark}

\author{P. M{\o}ller}
\affiliation{European Southern Observatory, Karl-Schwarzschildstrasse 2, 85748 Garching Bei Muenchen, Germany}

\author{M. A. Zwaan}
\affiliation{European Southern Observatory, Karl-Schwarzschildstrasse 2, 85748 Garching Bei Muenchen, Germany}

\author{J.~P.~U. Fynbo}
\affiliation{The Cosmic Dawn Center, Niels Bohr Institute, Copenhagen University, DK-2100 Copenhagen, Denmark}

\author{M. Dessauges-Zavadsky}
\affiliation{Observatoire de Gen{\`e}ve, Universit{\'e} de Gen{\`e}ve, 51 Ch. des Maillettes, 1290 Sauverny, Switzerland}


\begin{abstract}
We have used the Atacama Large Millimeter/submillimeter Array (ALMA) to carry out a search for CO (3$-$2) or 
(4$-$3) emission from the fields of 12 high-metallicity ([M/H]~$\geq -0.72$\,dex) damped Lyman-$\alpha$ absorbers 
(DLAs) at $z \approx 1.7-2.6$. We detected CO emission from galaxies in the fields of five DLAs (two of which 
have been reported earlier), obtaining high molecular gas masses, $\mmol \approx (1.3 - 20.7) 
\times (\alpha_{\rm CO}/4.36) \times 10^{10} \; \msun$. 
The impact parameters of the CO emitters to the QSO sightline lie in the range $b \approx 5.6-100$~kpc, with the three 
new CO detections having $b \lesssim 15$~kpc. The highest CO line luminosities and inferred molecular gas masses are
associated with the highest-metallicity DLAs, with [M/H]~$\gtrsim -0.3$\,dex. The high inferred molecular gas masses may 
be explained by a combination of a stellar mass-metallicity relation and a high molecular gas-to-stars mass ratio in 
high-redshift galaxies; the DLA galaxies identified by our CO searches have properties consistent with those of 
emission-selected samples. None of the DLA galaxies detected in CO emission were identified in earlier optical or 
near-IR searches and vice-versa; DLA galaxies earlier identified in optical/near-IR searches were not detected in 
CO emission. The high ALMA CO and C[{\sc ii}]~158$\mu$m detection rate in high-$z$, high-metallicity DLA galaxies 
has revolutionized the field, allowing the identification of dusty, massive galaxies associated with high-$z$ DLAs.
The \hi-absorption criterion identifying DLAs selects the entire high-$z$ galaxy population, including dusty 
and UV-bright galaxies, in a wide range of environments.

\end{abstract}


\section{Introduction} 
\label{sec:intro}

The most direct way of identifying galaxy populations at high redshifts is to detect the emission from 
individual galaxies in deep images, usually in the optical or the near-infrared bands.
However, such ``emission-selected'' samples contain a bias towards the more-luminous members of the 
population. Galaxies identified by their stellar emission have an additional bias towards objects with a 
high star-formation rate (SFR) and a high stellar mass. An alternative way of identifying high-$z$ galaxies, 
without luminosity or stellar biases, is via their damped Lyman-$\alpha$ absorption signature (with \hi\ column 
density $\geq 2 \times 10^{20}$~\cm) in quasar absorption spectra, if the galaxy lies along the sightline to 
a background quasar \citep[][]{wolfe05}. 
Quasar spectroscopy with the Sloan Digital Sky Survey (SDSS) has yielded 
more than 30,000 such damped absorbers (DLAs) today, mostly at $z > 2$ \citep[e.g.][]{noterdaeme12b}. 

Selected by the presence of high \hi\ column densities, absorption-selected galaxies provide a complementary view 
of the high-redshift universe to the usual emission-selected samples. Absorption-selected galaxies are not biased
towards high luminosities, SFRs, or stellar masses, although the absorption selection favours galaxies with larger gas
cross-sections. An important question in galaxy evolution is whether absorption and emission selection trace the same 
galaxy population, or whether new types of galaxies are identified by the absorption selection. Addressing this 
issue requires identification and characterization of the absorption-selected galaxies. Unfortunately, the presence of 
the nearby bright background quasar makes it difficult to even identify the DLA hosts via the usual techniques of optical 
imaging and spectroscopy \citep[although see][]{fumagalli14}. Only around 20 galaxies have been found by such studies 
to be associated with DLAs at $z \gtrsim 2$ \citep[e.g.][]{krogager17,mackenzie19}.
Further, DLA samples selected from magnitude-limited optical surveys like the SDSS are biased 
against dusty intervening galaxies, as a high dust content could obscure the background quasar and remove it from 
a magnitude-limited sample \citep[e.g.][]{krogager19}. DLA surveys towards radio- or mid-IR-selected quasar 
samples \citep[e.g.][]{ellison01,krogager16}, and ``blind'' \hii\ or mm-wave absorption surveys 
\citep[e.g.][]{kanekar14b} may be used to yield absorption-selected galaxies without a dust bias.

The advent of the Atacama Large Millimeter/submillimeter Array (ALMA) has allowed two new approaches to identify
and study galaxies associated with high-$z$ DLAs, tracing cold gas in emission with the redshifted \cii\ 
fine-structure and CO rotational lines. \citet{neeleman17,neeleman19} used ALMA \cii\ studies to identify the 
host galaxies of five high-metallicity DLAs at $z \approx 4$. Two of these galaxies were later mapped in their 
\cii\ emission, revealing a merging system at $z \approx 3.80$ \citep{prochaska19} and the ``Wolfe disk'',  
a cold rotating disk galaxy at $z \approx 4.26$ \citep{neeleman20}. ALMA CO studies of known DLA hosts at 
$z \approx 0.5-0.8$ yielded high molecular gas masses and large gas depletion times, very different 
from emission-selected samples \citep{moller18,kanekar18}. Our initial searches for CO emission associated 
with DLAs at $z \approx 2$ have identified massive galaxies in the fields of the $z = 2.193$ DLA towards 
B1228-113 \citep[][]{neeleman18} and the $z = 2.5832$ DLA towards J0918+1636 \citep{fynbo18}. 
And a Very Large Array (VLA) search for CO emission from the $z \approx 4.26$ \cii-emitting DLA host towards 
J0817+1351 also yielded a high molecular gas mass, $\approx 10^{11} \; \msun$ \citep{neeleman20}.

For a given molecular gas mass, it is easier to detect CO emission from high-metallicity galaxies than 
from low-metallicity ones, as the former have a lower CO-to-H$_2$ conversion factor $\aco$ \citep[e.g.][]{bolatto13}.
For example, in the case of normal (i.e. main-sequence) emission-selected galaxies at high redshifts, CO searches 
have typically been carried out in objects with a high stellar mass and/or a high SFR, consistent with a high 
metallicity \citep[e.g.][]{daddi10b,tacconi13,dessauges15,genzel17}. We hence chose to target high-metallicity 
absorbers in our search for CO emission from galaxies associated with high-$z$ DLAs. In this {\it Letter}, we report 
the results of an ALMA search for redshifted CO emission from the fields of 12~high-metallicity DLAs at 
$z \approx 2$.\footnote{We assume a flat $\Lambda$-Cold Dark Matter
cosmology, with $\Omega_\Lambda = 0.685$, $\Omega_m = 0.315$, H$_0 = 67.4$~\kms~Mpc$^{-1}$ \citep{planck18}.}

\section{Observations, Data analysis, and Results}

The ALMA bands~3 and 4 were used to search for redshifted CO emission from the fields of 12~DLAs at $z \approx 
1.66 - 2.58$ in proposals 2016.1.00628.S and 2017.1.01558.S (PI: Prochaska), from 2017~January to 2018~August.
The DLAs were selected to have high metallicities, [M/H]~$\geq -0.72$.
Four 1.875~GHz spectral windows were used for all observations, with one spectral window, sub-divided into 
3840~channels, used to cover either the CO($3-2$) or the 
CO($4-3$) line at the DLA redshift, and the remaining three spectral windows, sub-divided into 120~channels, used 
to measure the continuum emission in the field. The observations were carried out in compact array configurations,
with total on-source times of $0.2 - 2.4$~hours. A standard calibration approach was followed, with each observing 
run including observations of one or more flux calibrators and a bandpass calibrator, and with scans on the target 
source interleaved with scans on a secondary calibrator. The typical error on the flux density scale is expected to be $\lesssim 10$\%.

The initial calibration used the standard ALMA pipeline in the {\sc casa} package \citep{mcmullin07}. In 
three sources (B1228-113, B1230-101, J2225+0527), the quasar continuum flux density was sufficient to perform 
self-calibration; this was carried out in either the {\sc aips} \citep{greisen03} or the {\sc casa} packages,
following standard procedures. After subtracting out any detected continuum emission (using the {\sc casa} task 
{\sc uvsub}), the final spectral cubes were created using natural weighting, at velocity resolutions of 
$\approx 50-200$~\kms, and searched for line emission.  In case of non-detections, limits on the CO line flux 
density were obtained from the cubes at a velocity resolution of 200~\kms. 

\setcounter{table}{0}
\begin{table*}
\centering
\caption{Observations, data analysis, and results. The columns are (1)~the QSO name, (2)~the QSO redshift, (3)~the DLA 
redshift, $z_{\rm abs}$, (4)~the CO transition, (5)~the expected redshifted CO line frequency, in GHz, (6)~the RMS noise on the 
continuum image, in $\mu$Jy/Bm, (7)~the velocity resolution in \kms, of the final CO spectral cube, (8)~the RMS noise on 
the spectral cube at this velocity resolution, in mJy/Bm, (9)~the synthesized beam of the spectral cube, 
in $'' \times ''$, (10)~the integrated CO line flux density, in Jy~\kms, or the $3\sigma$ limit on this quantity, (11)~for CO detections, 
the peak redshift, from a single Gaussian fit, (12)~for CO detections, the line FWHM, $\rm W50_{CO}$, of the CO emission, in \kms, from the Gaussian
fit, (13)~the continuum flux density of the DLA galaxy, in $\mu$Jy, or the $3\sigma$ limit on this quantity.
\label{table:table1}}
\begin{tabular}{|c|c|c|c|c|c|c|c|c|c|c|c|c|}
\hline
QSO        & $z_{\rm QSO}$ & $z_{\rm abs}$ & CO line & $\nu_{\rm CO}$ & RMS & Resn. & RMS$_{\rm CO}$ & Beam & $\int S_{\rm CO} d{\rm V}$ & $z_{\rm CO}$ & $\rm W50_{CO}$ & $S_{\rm cont}$ \\
           &               &                   &         &    GHz         & $\mu$Jy/Bm           &   \kms         &  $\mu$Jy            & $'' \times ''$ &               Jy~\kms    &      & \kms  & $\mu$Jy \\
\hline
J0044+0018$^\star$   & 1.874 & 1.7250  &  $3-2$  & 126.9 &  16 & 200  & 125  & $3.2 \times 1.9$ & $<0.080         $ & $-$    & $-$   & $< 48$   \\
J0815+1037	         & 2.020 & 1.8462  &  $4-3$  & 162.0 &  11 & 200  &  80  & $2.2 \times 2.0$ & $< 0.051        $ & $-$    & $-$   & $< 33$   \\
J1024+0600           & 2.130 & 1.8950  &  $4-3$  & 159.3 &  19 & 200  & 157  & $2.2 \times 2.0$ & $<0.100         $ & $-$    & $-$   & $< 57$   \\
J2206-1958           & 2.558 & 1.9200  &  $4-3$  & 157.9 & 9.0 & 200  &  75  & $2.3 \times 1.6$ & $<0.048         $ & $-$    & $-$   & $< 27$   \\
B1230-101$^\star$    & 2.394 & 1.9314  &  $4-3$  & 157.3 &  23 & 200  &  90  & $2.3 \times 1.6$ & $<0.057         $ & $-$    & $-$   & $< 69$  \\
B0551-366            & 2.317 & 1.9622  &  $4-3$  & 155.6 &  24 &  50  & 330  & $2.5 \times 1.7$ & $0.433 \pm 0.052$ & $1.9615(2)$ & $300$ & $< 72$   \\
J0016-0012           & 2.090 & 1.9730  &  $4-3$  & 155.2 & 5.4 & 100  &  65  & $2.3 \times 1.8$ & $0.102 \pm 0.015$ & $1.9712(3)$ & $325$ & $< 16$   \\
J1305+0924           & 2.051 & 2.0184  &  $3-2$  & 114.6 &  13 & 200  & 200  & $3.1 \times 2.4$ & $<0.13          $ & $-$    & $-$   & $< 39$   \\
J2225+0527           & 2.323 & 2.1310  &  $3-2$  & 110.4 &  18 & 200  & 275  & $3.5 \times 2.7$ & $0.328 \pm 0.064$ & $2.1312(5)$ & $400$ & $-$      \\
B1228-113$^\dagger$  & 3.528 & 2.1929  &  $3-2$  & 108.3 & 11 & 43.2 & 150  & $2.6 \times 1.7$ & $0.726 \pm 0.031$ & $2.1933(5)$ & $600$ & $46 \pm 10$ \\
J2222-0946           & 2.926 & 2.3543  &  $3-2$  & 103.1 &  10 & 200  & 100  & $3.6 \times 2.2$ & $<0.064         $ & $-$    & $-$   & $< 30$ \\
J0918+1636$^\dagger$ & 3.096 & 2.5832  &  $3-2$  &  96.5 &  13 & 48.5 & 240  & $3.5 \times 2.6$ & $0.736 \pm 0.045$ & $2.5848(5)$ & $350$ & $40 \pm 13$ \\
\hline
\end{tabular}
\vskip 0.1in
$^\star$~Tentative ($\approx 4.5\sigma$ significance) detections of CO emission (listed as upper limits in the table).\\
$^\dagger$~The CO detections in the fields of B1228-113 and J0918+1636 were originally reported by \citet{neeleman18} and \citet{fynbo18}, respectively. The data were re-analysed here, for consistency (yielding lower errors on the integrated CO line flux density).\\
\end{table*}

\begin{figure*}[t!]
\centering
\includegraphics[width=2.2in]{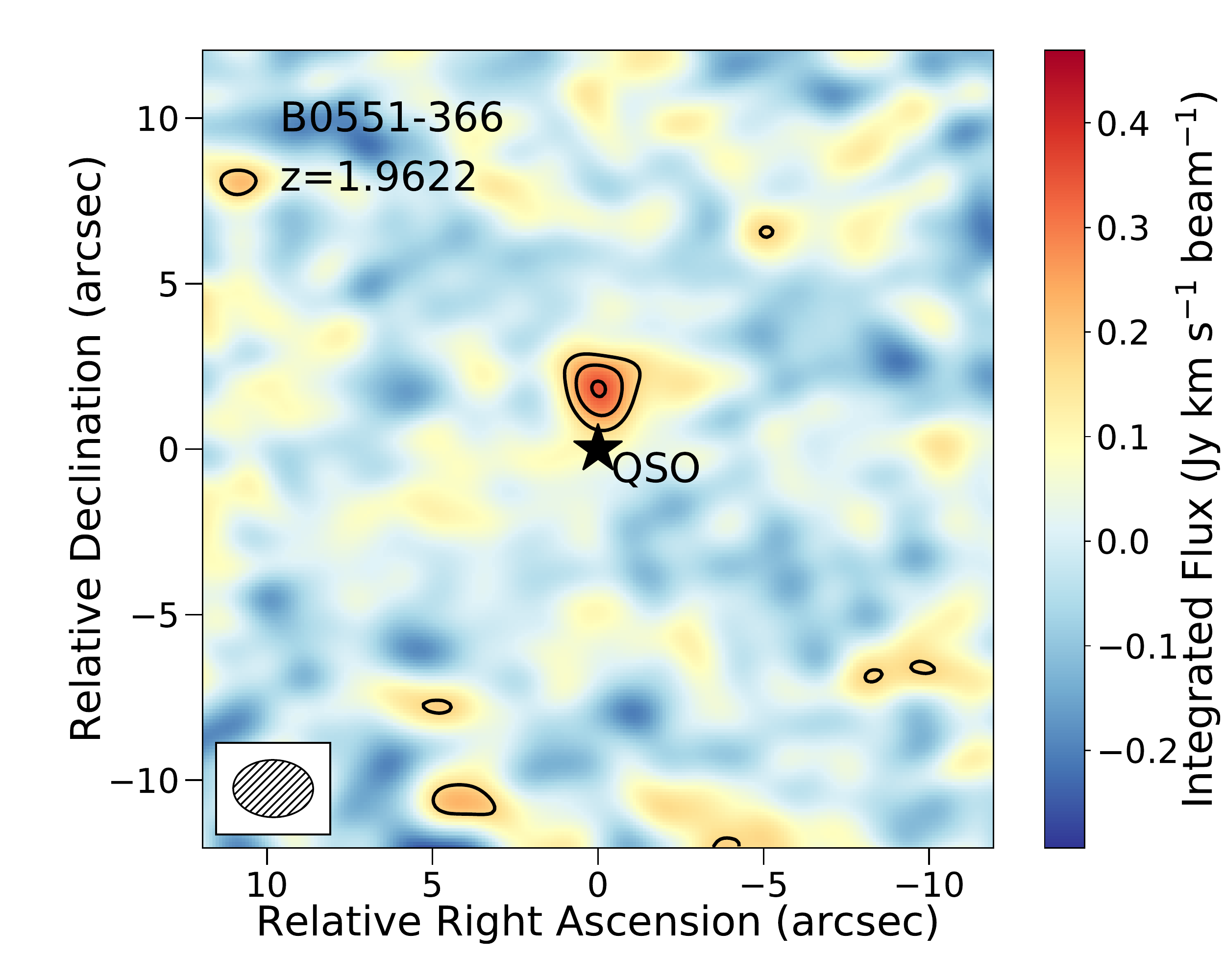}
\includegraphics[width=2.2in]{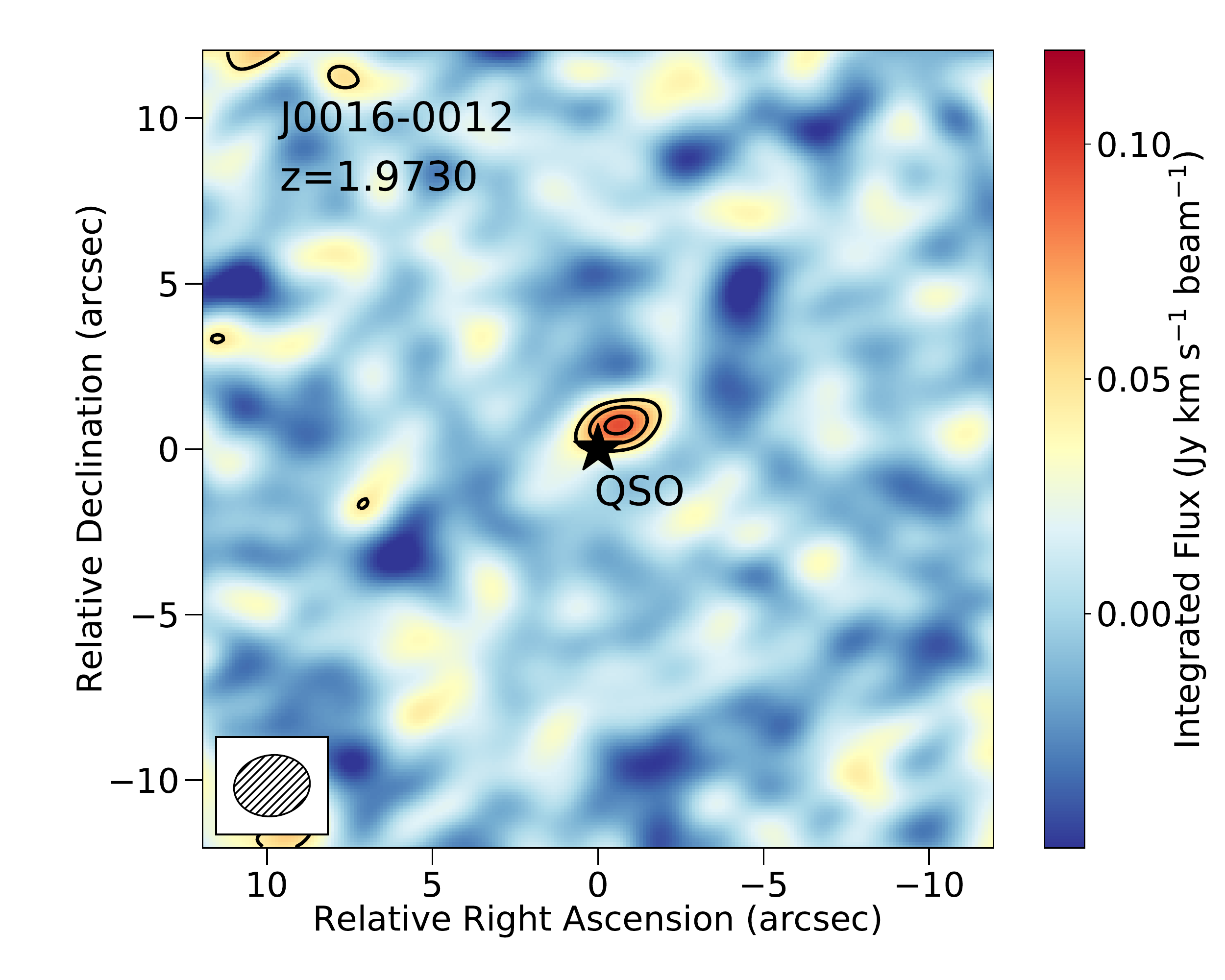}
\includegraphics[width=2.2in]{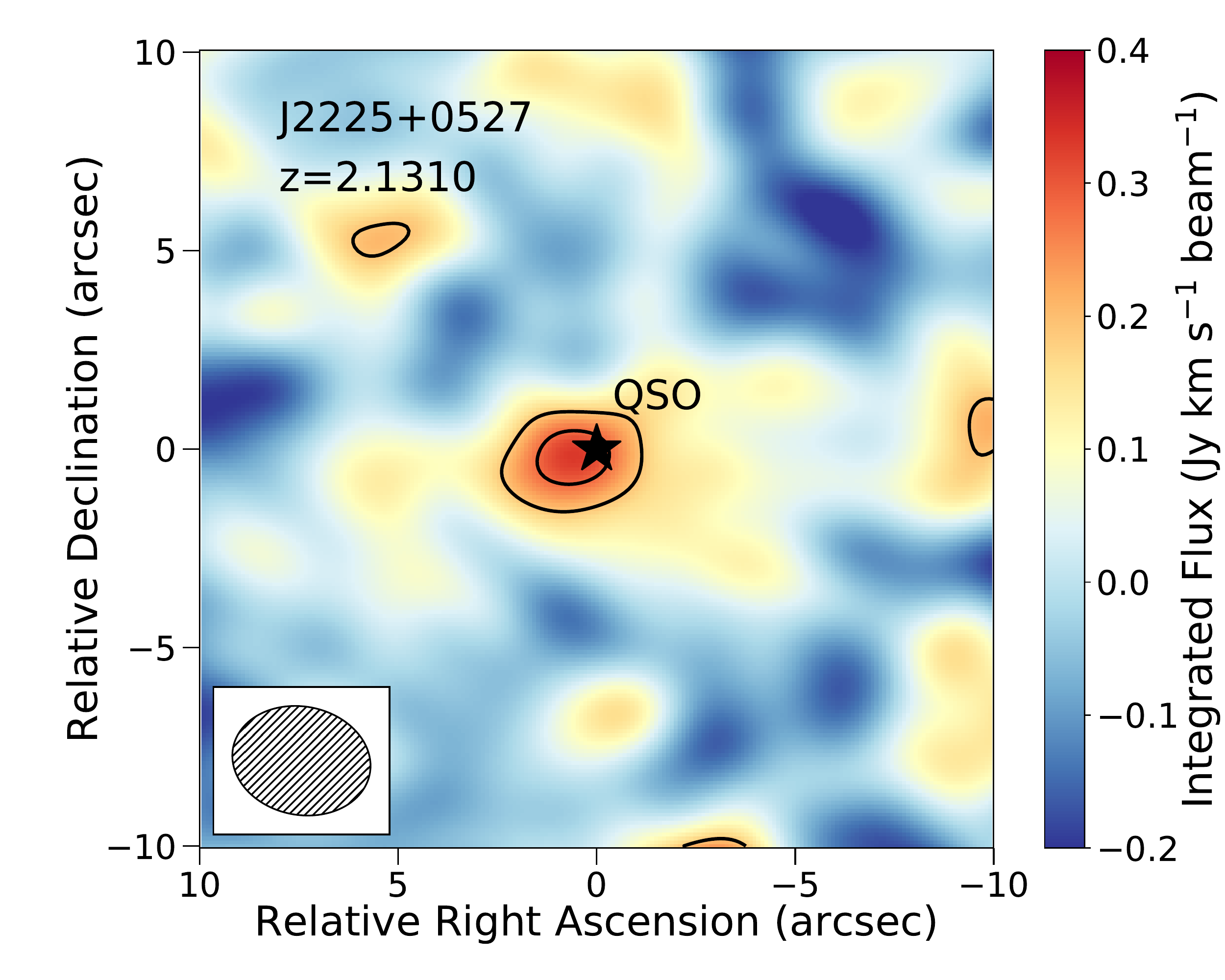}
\includegraphics[width=2.2in]{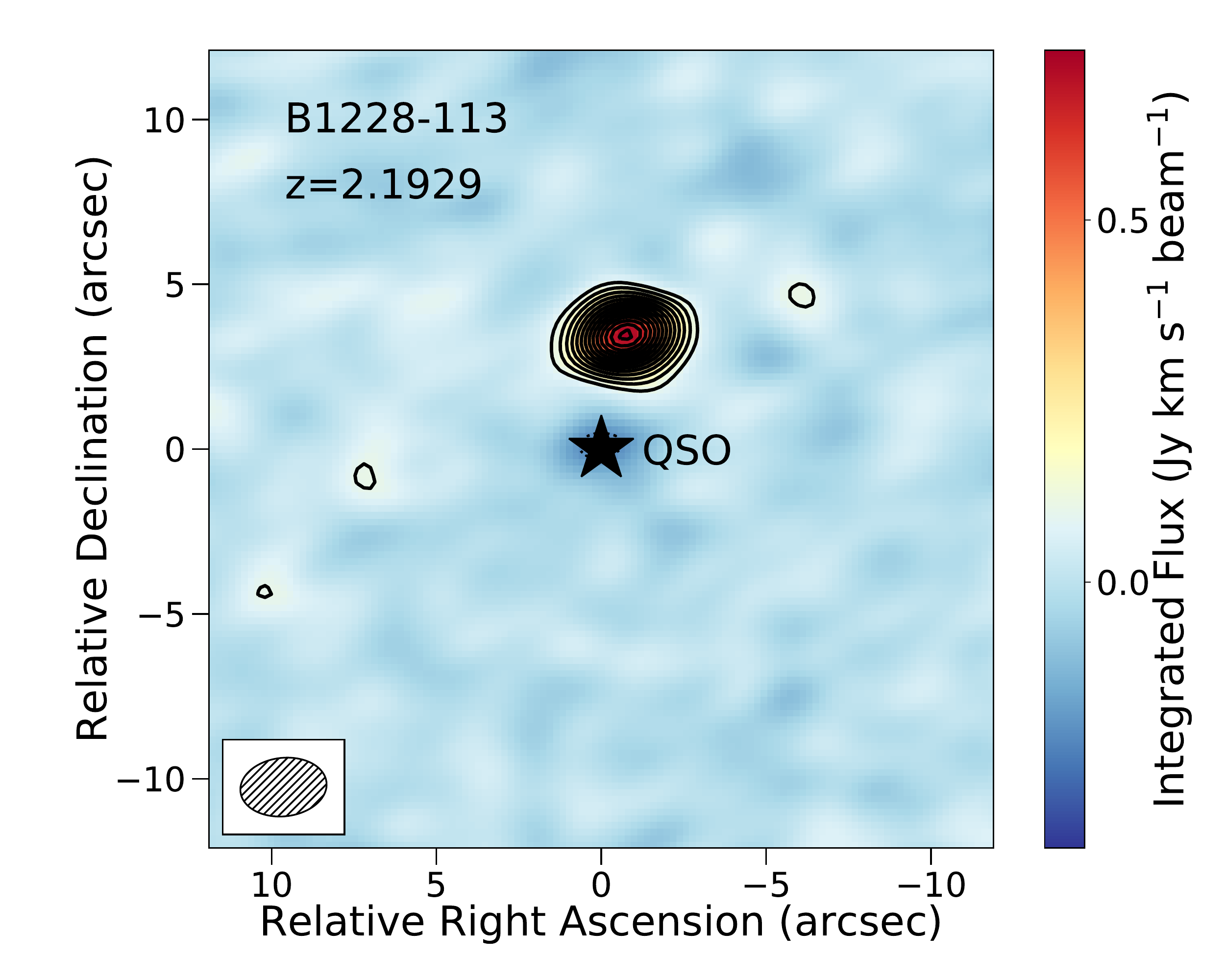}
\includegraphics[width=2.2in]{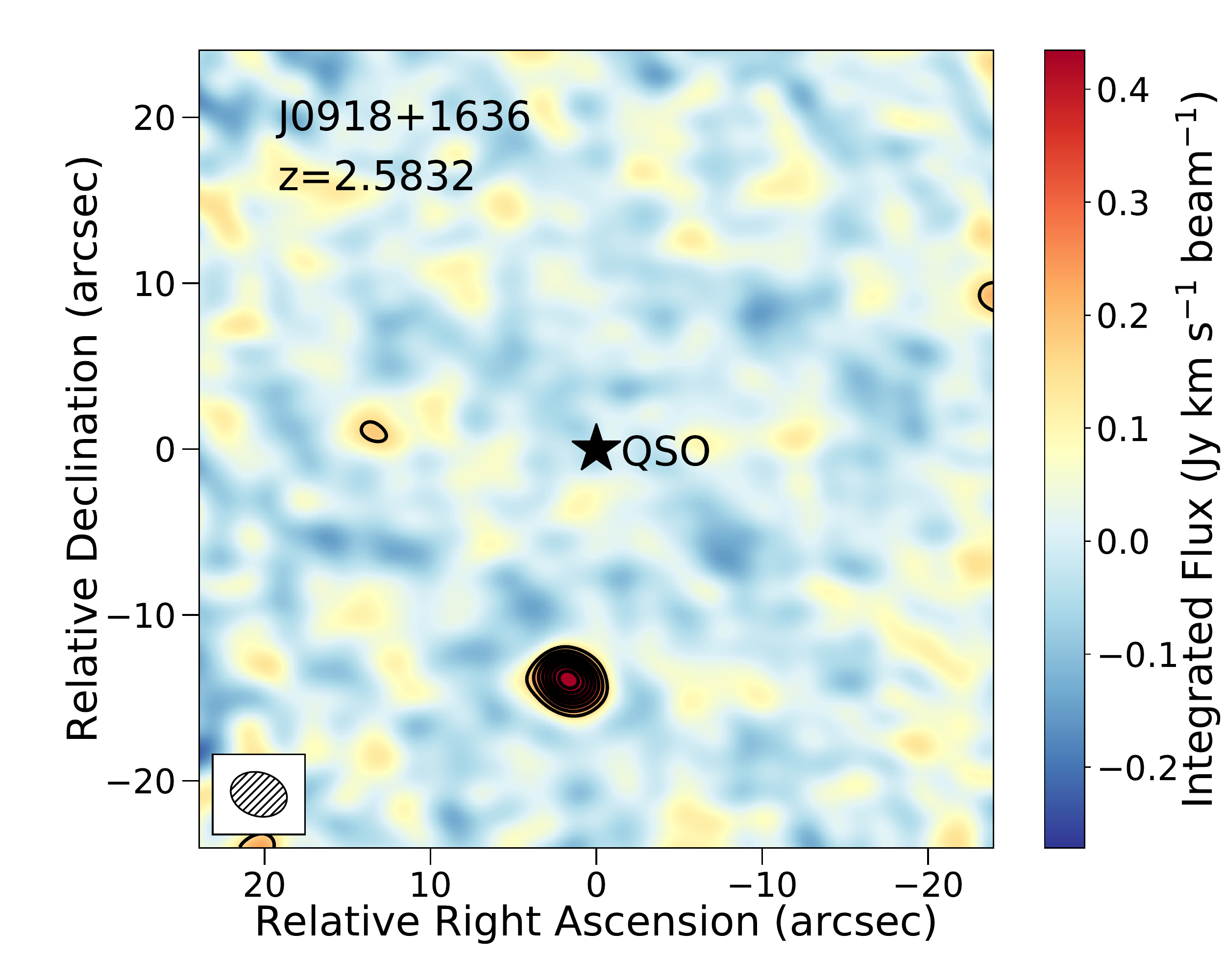}
\caption{Velocity-integrated CO emission for the five CO detections, in order of increasing redshift.
The quasar name and the DLA redshift are indicated at the top left corner of each panel, and the axes 
co-ordinates are relative to the quasar's J2000 co-ordinates. The solid contours are at 
$(3,4.2,6, ...)$~$\times \sigma$, and the negative (dashed) contour at $-3\sigma$, where $\sigma$ is the RMS 
noise on each image. The hatched ellipse at the bottom left of each panel shows the ALMA synthesized beam.
\label{fig:fig1}}
\end{figure*}

\begin{figure*}[t!]
\centering
\includegraphics[width=2.2in]{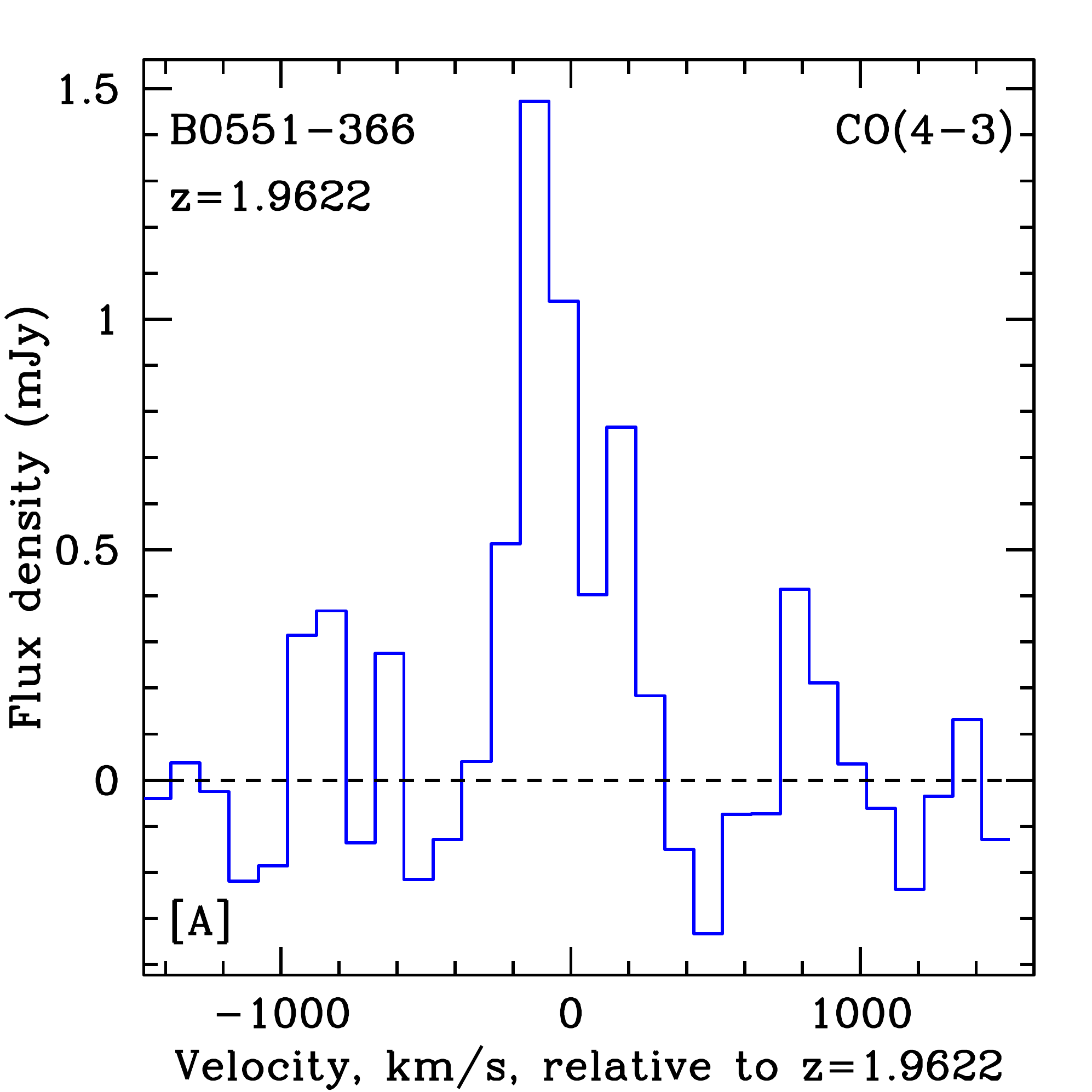}
\includegraphics[width=2.2in]{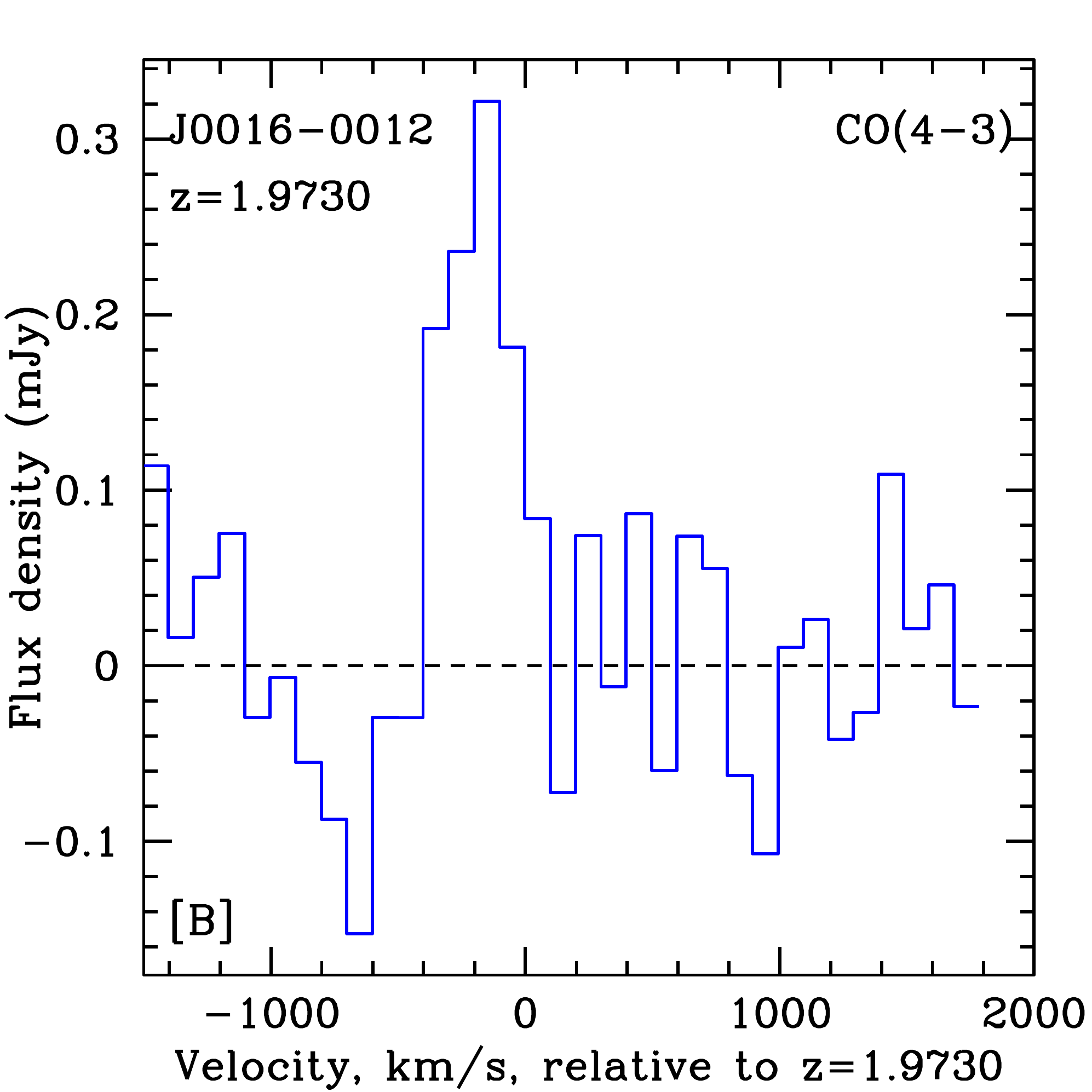}
\includegraphics[width=2.2in]{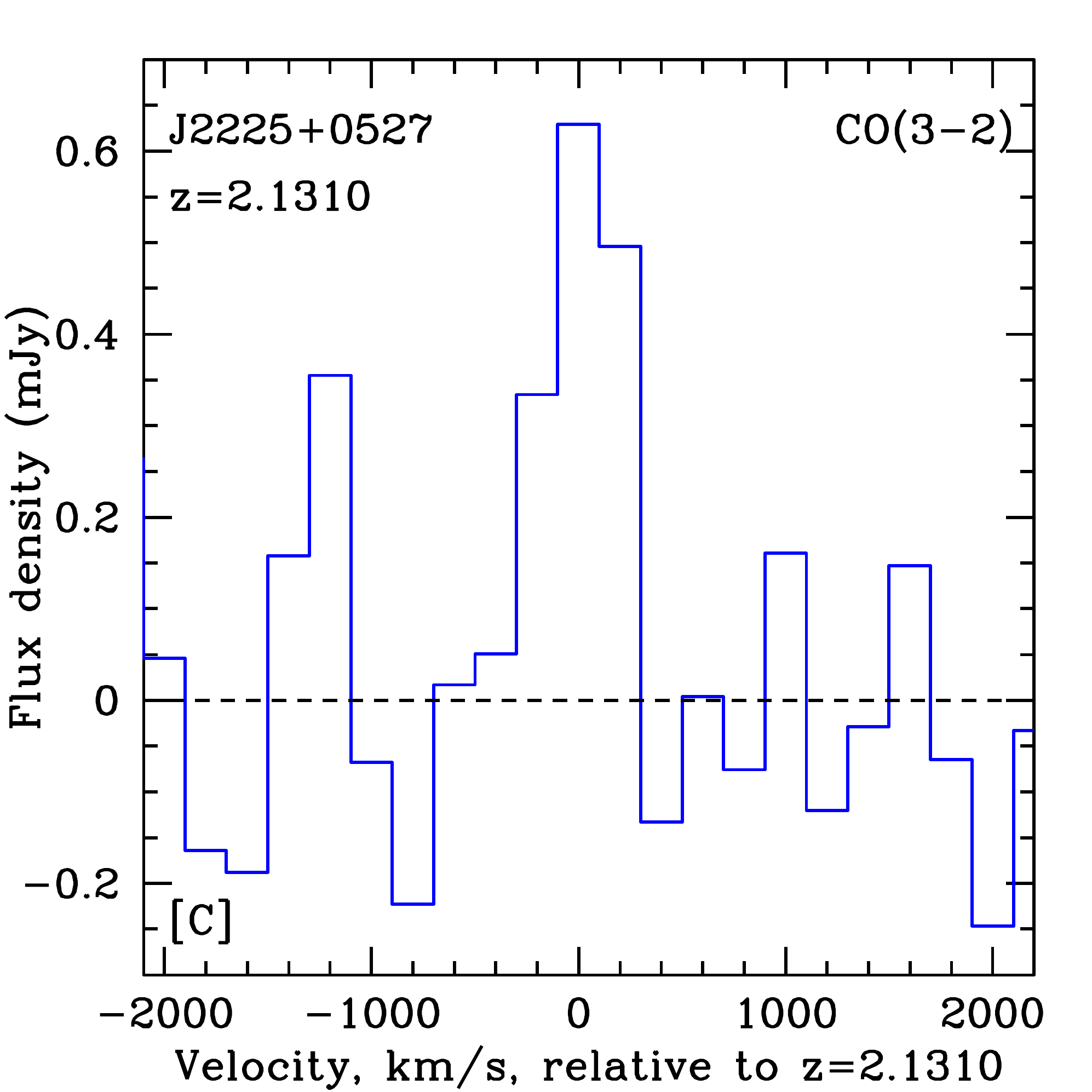}
\includegraphics[width=2.2in]{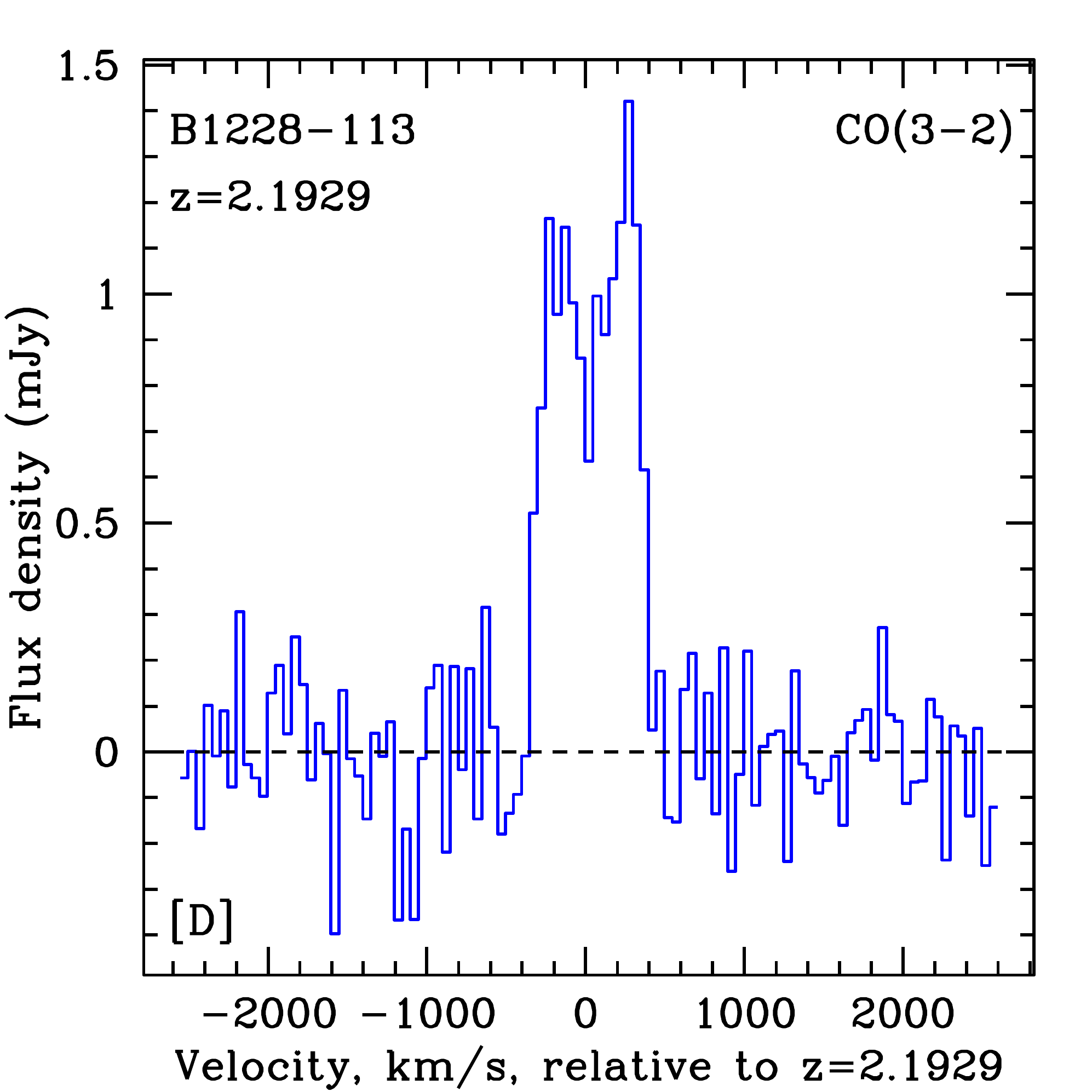}
\includegraphics[width=2.2in]{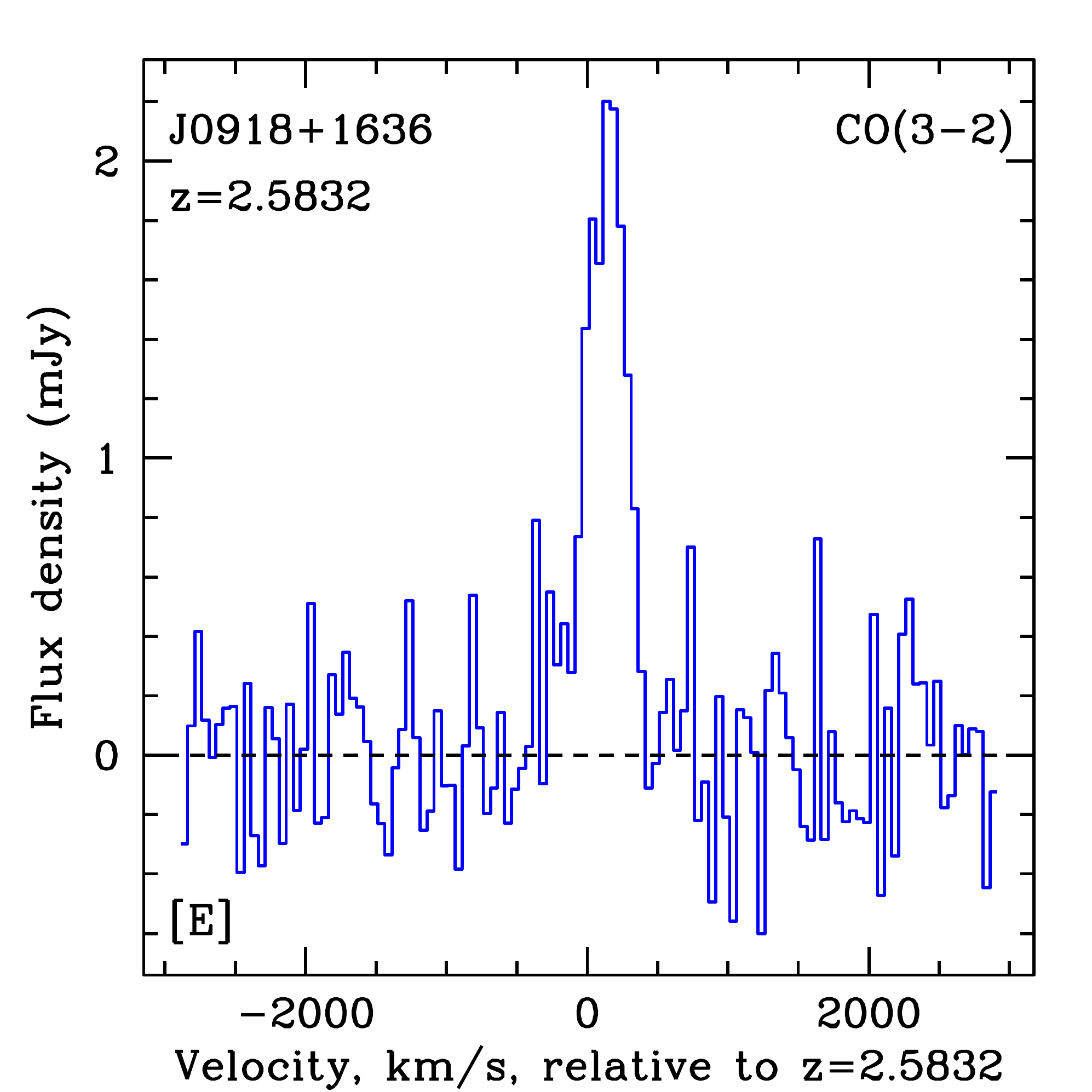}
\caption{CO spectra for the five detections:
[A]~B0551-366, at $100$~\kms\ resolution, [B]~J0016-0012, at $100$~\kms\ resolution, 
[C]~J2225+0527, at $200$~\kms\ resolution, [D]~B1228-113, at $50$~\kms\ resolution, and [E]~J0918+1636, at $50$~\kms\ resolution.
\label{fig:fig2}}
\end{figure*}

Line emission was clearly detected (at $> 5\sigma$ significance) from galaxies in five DLA fields, 
B0551-366 at $z = 1.9622$, J0016-0012 at $z=1.9730$, J2225+0527 at $z = 2.1310$, B1228-113 at $z = 2.1929$, 
and J0918+1636 at $z = 2.5832$, at velocities in good agreement with those of the low-ionization metal absorption 
lines.  Fig.~\ref{fig:fig1} shows the velocity-integrated CO emission from the five  CO detections, two of which 
(in the fields of B1228-113 and J0918+1636) have been presented earlier \citep{neeleman18,fynbo18} 
In all cases, the emission is unresolved by the ALMA synthesized beam; the final CO spectrum was hence obtained from a cut 
through the ``dirty'' cube at the location of the peak pixel of the velocity-integrated CO emission. In passing, we note that 
the ALMA synthesized beam subtends a physical size $\gtrsim 15$~kpc at the DLA redshift for all five galaxies; it is thus 
unlikely that we are resolving out any CO emission. For comparison, \citet{daddi10b} obtain CO sizes of $\approx 6-11$~kpc in 
their sample of $z\approx 1.5$ BzK galaxies. The final CO line spectra of the five detections are shown in 
Fig.~\ref{fig:fig2}.

Two DLA fields,  J0044+0018 at $z = 1.7250$ and B1230-101 at $z = 1.9314$, showed tentative ($\approx 4.5-4.7\sigma$)
detections of line emission, but $\approx 300-400$~\kms\ away from the absorption redshift. We note that 
\citet[][see their Table~1]{moller20} obtain a velocity offset $\leq 210$~\kms\ between the optical emission redshift and 
the absorption redshift in  ten DLAs at $z \approx 2$. If the tentative CO detections are confirmed, the velocity 
offset between the absorption and emission redshifts suggests that either the DLA arises in gas clumps in the 
circumgalactic medium of the CO emitters, or the CO emission arises from galaxies in the vicinity of the DLA host, 
but not from the absorbing galaxy itself \citep[as is the case for the CO emitter towards J0918+1636;][]{fynbo18}.

The details of the observations, data analysis, and results are summarized in Table~\ref{table:table1}, in order 
of increasing DLA redshift. The penultimate column lists the full-width-at-half-maximum (FWHM) of the CO emission 
for the five detections; the line FWHMs are all large, $\approx 325-600$~\kms, consistent with massive galaxies, 
with the kinematics likely to be dominated by either rotation or mergers. The last column lists the continuum flux density of the 
identified DLA galaxy at the CO rest frequency, or $3\sigma$ upper limits on this quantity. 
Finally, for J2225+0527, where the continuum emission from the DLA galaxy is unresolved from the strong QSO 
continuum emission, no value is listed here.

The measured CO($3-2$) or CO($4-3$) line flux densities may be used to infer the CO line luminosities, L$'_{{\rm CO}(3-2)}$
or L$'_{{\rm CO}(4-3)}$ \citep{carilli13}, and thence, the molecular gas mass in the DLA galaxies, if one knows the 
CO-to-H$_2$ conversion factor $\aco$ and the CO level excitation \citep{carilli13,bolatto13}. We will assume sub-thermal 
excitation of the higher-J levels, with $R_{13} \approx 1.8$ and $R_{14} \approx 2.4$ 
[where $R_{13} = {\rm L}'_{{\rm CO}(1-0)}/{\rm L}'_{{\rm CO}(3-2)}$ and $R_{14} = {\rm L}'_{{\rm CO}(1-0)}/{\rm L}'_{{\rm CO}(4-3)}$], 
valid for galaxies near the main sequence at $z \approx 0-3$ \citep[e.g.][]{tacconi20}. We will also assume $\aco = 4.36 \; 
\msun$~(K~\kms~pc$^{2}$)$^{-1}$, applicable for galaxies with metallicity near solar and which are not undergoing a starburst 
\citep[e.g.][]{bolatto13,tacconi20}. The inferred molecular gas masses for the five DLA galaxies with CO detections lie in the range 
$(1.3 - 20.7) \times (\aco/4.36) \times 10^{10} \; \msun$, while the $3\sigma$ upper limits on the molecular gas mass for the CO 
non-detections are $(0.6 - 2.3) \times (\aco/4.36) 10^{10} \; \msun$. While a lower value of $\aco$ or a higher CO line
excitation \citep[e.g.][]{klitsch19,riechers20} would lower the above molecular gas mass estimates, we note that the assumed values 
are likely to be reliable for normal (i.e. non-starburst) galaxies. Our results are summarized in Table~\ref{table:results}.

\newpage
\section{Discussion}
\label{sec:discuss} 

Our ALMA search for CO emission from the galaxies associated with 12 high-metallicity DLAs at $z \approx 2$ has 
resulted in the confirmed ($> 5 \sigma$ significance) detection of five CO-emitting galaxies at redshifts 
consistent with the DLA redshift \citep[see also][]{neeleman18,fynbo18}.
Along with the detection of CO(2-1) emission from the Wolfe disk \citep{neeleman20}, these are the first detections 
of CO emission from high-$z$ DLA galaxies, after more than 25 years of such searches \citep[e.g.][]{wiklind94b}.
ALMA has revolutionized this field, providing a new route to identifying DLA galaxies unaffected by the QSO's brightness.

\begin{figure*}[t!]
\centering
\includegraphics[width=3.0in]{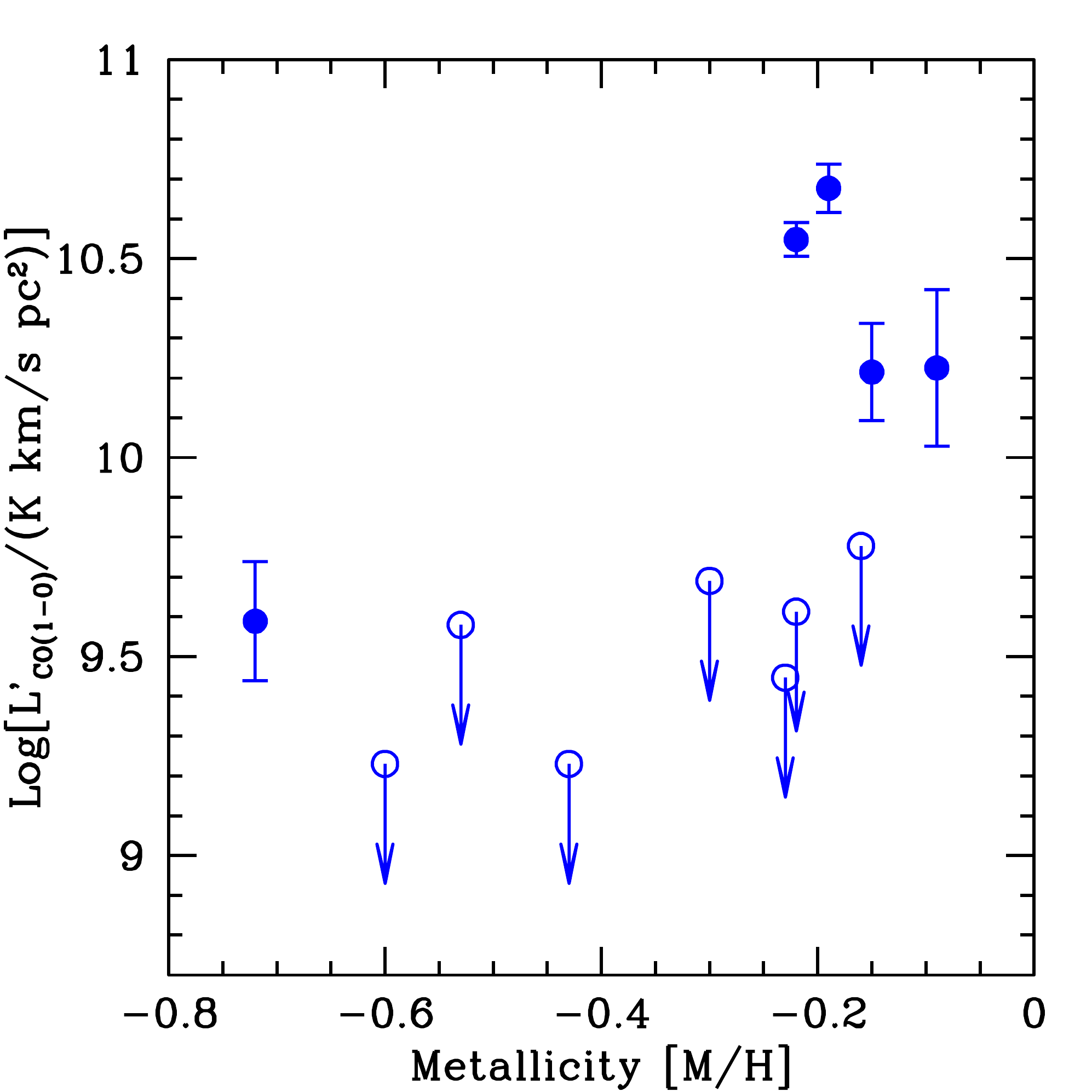}
\includegraphics[width=3.0in]{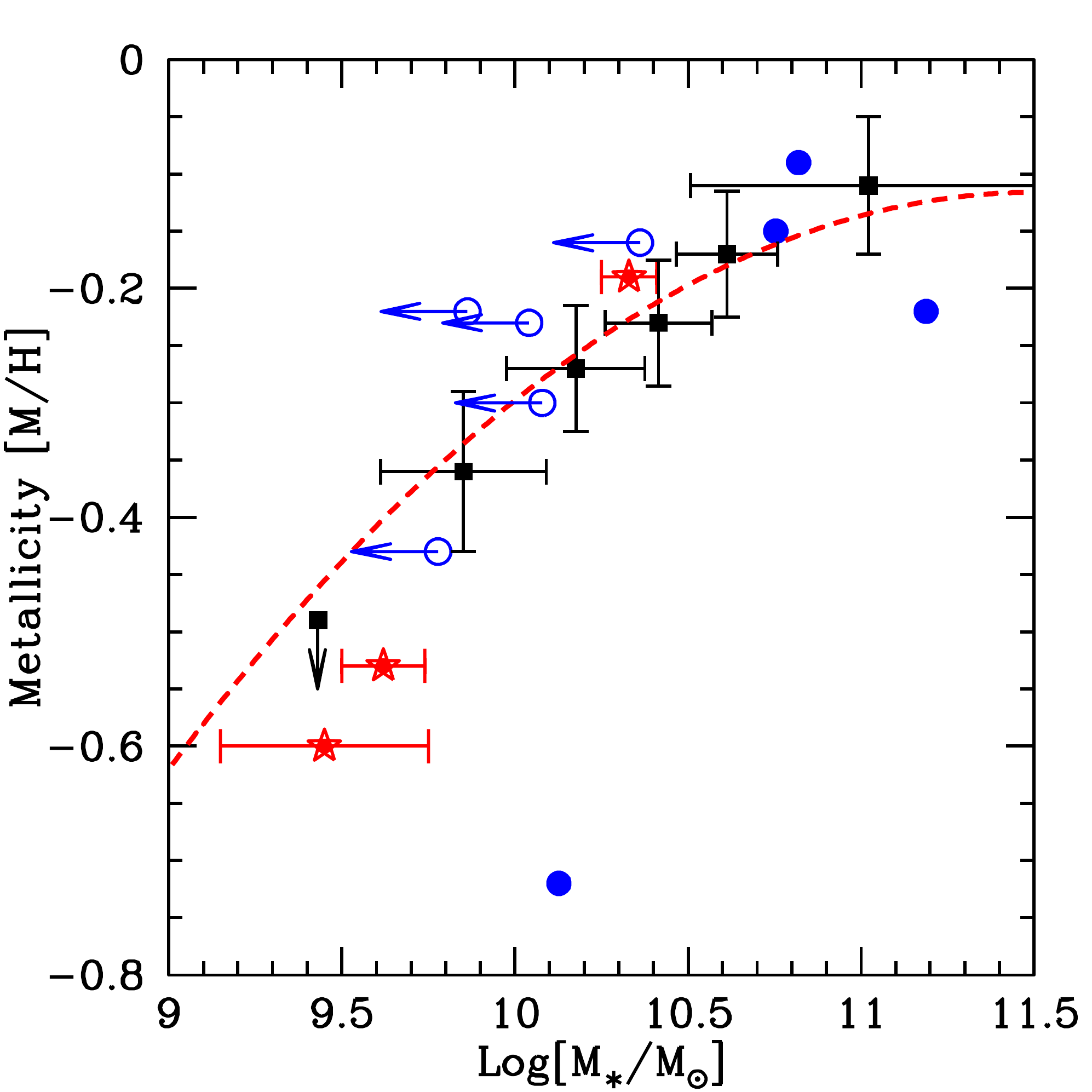}
\caption{[A]~(left panel): The logarithim of the inferred CO(1--0) line luminosity, $\rm Log[L'_{CO(1-0)}/(K~km s^{-1} pc^2)]$, 
plotted against absorption metallicity [M/H]; the highest values of $\rm L'_{CO(1-0)}$ are seen to be obtained at the highest 
values of [M/H]. [B]~(right panel): The absorption metallicity [M/H] plotted against the logarithm of the inferred 
stellar mass (assumed to be equal to the molecular gas mass), with CO detections (non-detections) shown as filled (open) 
blue circles. The filled black squares show the (binned) emission metallicity plotted against the (binned) stellar mass 
for the UV-selected galaxies of \citet{erb06}. The dashed red curve shows the mass-metallicity relation of nearby galaxies
\citep{tremonti04}, shifted down by 0.56~dex \citep{erb06}. The three DLA galaxies identified via optical spectroscopy, 
J2222-0946, J2206-1958, and J0918+1636 \citep{fynbo13,krogager13,christensen14}, are shown as red stars, with stellar mass
estimates from the optical/near-IR photometry.
\label{fig:mstar-met}}
\end{figure*}

The DLAs of the sample have [M/H]~$\geq -0.72$, and thus lie in the top quartile of the metallicity distribution of 
DLAs at $z \approx 2$ \citep[e.g.][]{rafelski14}. Athough the sample size is still small, the CO detections tend to 
arise (see Table~\ref{table:results}) in the DLAs with the highest metallicities and $\Delta V_{\rm 90}$ values, 
of our sample. For example, four of the five CO detections are in the fields of DLAs with [M/H]~$> -0.3$; these four 
galaxies have the highest inferred CO(1--0) line luminosities ($> 10^{10}$~K~\kms~pc$^2$; see Fig.~\ref{fig:mstar-met}[A]) 
and the highest inferred molecular gas masses, $\gtrsim 6 \times 10^{10} \; \msun$. Similarly, all five CO detections 
have $\Delta V_{\rm 90} > 160$~\kms. Earlier studies have found such high $\Delta V_{\rm 90}$ values to be associated 
with massive DLA galaxies at small impact parameters to the QSO sightline \citep{christensen19,moller20}; our five 
detections of CO emission do indeed arise in massive galaxies, but three of the 
impact parameters are large, $\approx 15-100$~kpc. Finally, the large FWHMs of the detected CO lines ($\approx 325-600$~\kms) 
are suggestive of rotation; for the $z = 2.1929$ DLA galaxy towards B1228-113, this is supported by the ``double-horned'' 
nature of the CO line profile (see Fig.~\ref{fig:fig2}[D]). However, our present ALMA data do not have the resolution or 
sensitivity to rule out the possibility of merging galaxies, as has been shown to be the case for the $z \approx 3.7978$ 
DLA towards J1201+2117 \citep{prochaska19}. 

Interestingly, we have detected CO emission from a galaxy in the field of the $z=1.9730$ DLA towards J0016-0012, which has 
the lowest metallicity ([M/H]~$=-0.72$; \citealp{petitjean02}) of our sample. The low impact parameter of the DLA galaxy to 
the QSO sightline ($\approx 6.7$~kpc) makes it unlikely that the emission metallicity is near solar, unless the galaxy has 
a large metallicity gradient ($\approx 0.12$~dex per kpc). If the metallicity of the DLA galaxy is below solar, it is 
likely that $\aco > 4.36 \; \msun$~(K~\kms~pc$^{2}$)$^{-1}$; if so, our estimate of the molecular gas mass in this galaxy 
would be a lower limit. However, this DLA also has by far the largest $\Delta V_{\rm 90}$ of our sample, $\approx 720$~\kms\ 
\citep{ledoux06}, and is the only system for which the $\Delta V_{\rm 90}$ value is larger than the width of the CO emission 
\citep[see also][]{neeleman17,prochaska19}. It is plausible that the large $\Delta V_{\rm 90}$ value is due to the presence of 
two or more galaxies along the QSO sightline \citep[see also][]{petitjean02}, both contributing to the \hi\ column density, but 
only one of which has a high (near-solar) metallicity and is detected in CO emission. Consistent with this hypothesis, strong 
metal-line absorption (including C{\sc i} and C{\sc i}$^*$, but, curiously, not H$_2$) was detected at $z = 1.9714$ by 
\citet{petitjean02}, in excellent agreement with the peak CO redshift ($z = 1.97120 \pm 0.00025$). In passing, we note that 
\citet{neeleman17} find the absorption spread $\Delta V_{\rm 90}$ of the $z = 3.7978$ DLA towards J1201+2117 to be larger than 
the width of the \cii\ emission of the associated galaxy; the \cii\ emission was later shown to arise from two merging galaxies
\citep{prochaska19}.

\begin{table*}[t!]
\centering
\caption{Derived properties of the DLAs of the sample and their associated galaxies. The columns are (1)~the QSO
name, (2)~the DLA redshift, (3)~the \hi\ column density, (4)~the gas-phase metallicity [M/H], using Zn, S, or Si as the tracer
\citep[e.g.][]{prochaska07,berg15}, (5)~the velocity width $\Delta V_{\rm 90}$ of the low-ionization metal absorption 
lines \citep[e.g.][]{prochaska97}, (6)~the CO transition, (7)~the CO($3-2$) or CO($4-3$) line luminosity (or the $3\sigma$ upper 
limit on this quantity), L$'_{\rm CO}$, (8)~the molecular gas mass (or the $3\sigma$ upper limit on the mass), $\mmol $, assuming 
$R_{13} = 1.8$ or $R_{14} = 2.4$, and $\aco = 4.36 \; \msun$~(K~\kms~pc$^{2}$)$^{-1}$, (9)~for CO detections, the impact 
parameter $b$, in kpc, between the QSO sightline and the CO emission, and (10)~references for the absorption data.
\label{table:results}}
\begin{tabular}{|c|c|c|c|c|c|c|c|c|c|c|}
\hline
DLA          & $z_{\rm abs}$ & $\nhi$ & [M/H] & $\Delta V_{\rm 90}$ & CO line & ${\rm L}'_{\rm CO}$ & $\mmol$ & $b$ & Refs. \\
             &               &  \cm   &       &   \kms              &         & $10^9$~K~\kms~pc$^2$ & $10^{10} \; \msun$ & kpc &  \\
\hline
J0044+0018$^\star$   & 1.7250 & 20.35 & -0.23 & 172 & $3-2$ & $<1.4         $  & $<1.1 $         & $-$    & 1 \\
J0815+1037	         & 1.8462 & 20.30 & -0.43 & $-$ & $4-3$ & $< 0.57       $  & $<0.60$         & $-$    & 1 \\
J1024+0600           & 1.8950 & 20.60 & -0.30 & 161 & $4-3$ & $<1.2         $  & $<1.2 $         & $-$    & 1 \\
J2206-1958           & 1.9200 & 20.67 & -0.60 & 136 & $4-3$ & $<0.57        $  & $<0.60$         & $-$    & 2 \\
B1230-101$^\star$    & 1.9314 & 20.48 & -0.22 & 94  & $4-3$ & $<0.70        $  & $<0.73$         & $-$    & 3 \\
B0551-366            & 1.9622 & 20.50 & -0.15 & 239 & $4-3$ & $5.42 \pm 0.65$  & $5.67 \pm 0.68$ & $15.4$ & 2 \\
J0016-0012           & 1.9730 & 20.83 & -0.72 & 720 & $4-3$ & $1.28 \pm 0.19$  & $1.34 \pm 0.20$ & $6.7$  & 4 \\
J1305+0924           & 2.0184 & 20.40 & -0.16 & 135 & $3-2$ & $<3.0         $  & $<2.3 $         & $-$    & 1 \\
J2225+0527           & 2.1310 & 20.69 & -0.09 & 331 & $3-2$ & $8.42 \pm 1.64$  & $6.6  \pm 1.3 $ & $5.6$  & 3 \\
B1228-113            & 2.1929 & 20.60 & -0.22 & 163 & $3-2$ & $19.63 \pm 0.84$ & $15.41 \pm 0.66$ & $30.0$ & 5 \\
J2222-0946           & 2.3543 & 20.65 & -0.53 & 174 & $3-2$ & $<1.9         $  & $<1.3 $         & $-$    & 6 \\
J0918+1636           & 2.5832 & 20.96 & -0.19 & 288 & $3-2$ & $26.4 \pm 1.6 $  & $20.7 \pm 1.3$  & $100$  & 7 \\
\hline
\end{tabular}
\vskip 0.1in
$^\star$~Tentative ($\approx 4.5\sigma$ significance) detections of CO emission.\\
References: (1)~\citet{berg15}; (2)~\citet{neeleman13}, (3)~\citet{krogager16}, (4)~\citet{petitjean02}, (5)~\citet{neeleman18}, 
(6)~\citet{fynbo10}, (7)~\citet{fynbo13}.
\vskip 0.1in
\end{table*}

Besides the high CO detection rate, our most striking result is the large molecular gas masses of the five galaxies with 
confirmed CO detections, $\mmol \approx (1.3 - 20.7) \times (\aco/4.36) \times 10^{10} \; \msun$. These are similar to 
the masses of the most massive colour-selected BzK galaxies at $z \approx 1.5-2.5$ \citep[e.g.][]{daddi10b}, and galaxies 
selected based on a combination of high stellar mass and high SFR at $z \approx 2.2$ \citep[e.g.][]{tacconi13}. The high 
molecular gas masses of a significant fraction of galaxies in the fields of high-metallicity DLAs at $z \approx 2$ can 
be plausibly accounted for by a combination of a mass-metallicity relation and a high molecular gas-to-stars mass ratio in 
high-$z$ galaxies. For the former, the emission metallicities of UV-selected galaxies are known to correlate with their 
stellar masses ($\mstar$) at high redshifts \citep{erb06}. Indeed, the metallicities and velocity spreads of low-ionization 
metal absorption lines in high-$z$ DLAs are consistent with the existence of a similar mass-metallicity relation in the 
associated galaxies \citep[e.g.][]{wolfe98,ledoux06,prochaska08}, as well as redshift evolution in this relation 
\citep{ledoux06,moller13,neeleman13}. For the latter, high-$z$ galaxies with high stellar masses and high SFRs have been shown 
to have a high molecular gas-to-stars mass ratio, $\mmol/\mstar \approx 1$ \citep[e.g.][]{dessauges15,tacconi20}. 

Fig.~\ref{fig:mstar-met}[B] plots the DLA metallicity versus the inferred stellar mass for the DLA galaxies, assuming 
$\mmol/\mstar \approx 1$. Also plotted in the figure are the emission metallicity (relative to solar, assuming a 
solar metallicity of 12+[O/H]=8.69; \citealp{asplund09})
and the stellar mass for the (binned) UV-selected galaxies of \citet{erb06}. It should be emphasized that the emission 
metallicities of the DLA galaxies are likely to be equal to or higher than the absorption metallicities \citep[e.g.][]{moller13}, 
given that the absorption metallicity is being measured in the outskirts of the galaxy (impact parameters 
$\approx 6-100$~kpc). Despite this, and the simplistic assumption that $\mstar \approx \mmol$, all but one of the 
absorption-selected DLA galaxies appear broadly consistent with the stellar mass-metallicity relation of 
emission-selected galaxies at similar redshifts. In passing, we emphasize that the fact that the DLA galaxies
appear to lie on the same mass-metallicity relation as the emission-selected galaxies does not necessarily imply 
that the absorption arises from gas in the specific DLA galaxy identified here (see below).

The three new CO detections, towards B0551-366, J0016-0012, and J2225+0527, all have relatively low impact parameters 
($b \approx 6-15$~kpc) to the QSO sightline; it is plausible that the damped Lyman-$\alpha$ absorption here arises from 
gas in the disks of these galaxies \citep[although the disk may be only one of the absorbing components along the sightline; 
see][]{christensen19,moller20}. This is especially the case for the DLAs towards J0016-0012 and J2225+0527, where 
$b \approx 5.6, 6.7$~kpc. However, we note that the peak CO emission redshift for three of the four CO-emitting galaxies 
with $b \lesssim 30$~kpc is in good agreement (within $\approx 75$~\kms) with the DLA redshift, while even for the fourth 
system, the DLA towards J0016-0012, strong low-ionization metal-line absorption is detected at $z = 1.9714$, consistent 
with the CO emission redshift. The agreement between the emission and absorption redshifts is surprising, given the wide 
range of impact parameters, $\approx 5.6-30$~kpc, and the large CO FWHMs and $\Delta V_{\rm 90}$ values (which indicate that 
the galaxies are unlikely to be close to face-on). This suggests that some of these DLAs may arise in gas clumps in a 
dynamically well-mixed CGM around the CO-emitting galaxies \citep[see also][]{christensen19,moller20}. Finally, the CO 
emission in the field of J0918+1636 has a large impact parameter ($b \approx 100$~kpc), a significant velocity offset 
($\approx 130$~\kms) between the low-ionization metal absorption lines and the CO emission line, as well as a galaxy at 
the DLA redshift much closer to the QSO sightline; this indicates that the CO emission here is likely to arise from a 
companion galaxy within the group of galaxies at the DLA redshift \citep{fynbo13,fynbo18}.

Five of the DLA fields of our sample (towards B1228-113, J2206-1958, J2222-0946, J2225+0527, and J0918+1636) have 
been earlier searched for the associated galaxies using optical or near-IR spectroscopy 
\citep[e.g.][]{moller02,peroux12,fynbo13,krogager16}. In the three cases where the associated galaxy has been confirmed 
by optical or near-IR spectroscopy (J2206-1958, J2222-0946, and J0918+1636), we did not obtain a detection of CO emission 
from the galaxy with the optical emission \citep[in the case of J0918+1636, we detected a second galaxy at the DLA 
redshift;][]{fynbo18}. Conversely, the three galaxies that showed CO emission (B1228-113, J2225+0527, and J0918+1636) 
were not identified in the optical or near-IR spectroscopy: the galaxy in the field of B1228-113 was found to show weak H$\alpha$ emission {\it after} its CO position was known \citep{neeleman18}, while the galaxy in the field of J2225+0527 was earlier tentatively detected in a near-IR image \citep{krogager16}, but without a confirming redshift. This suggests that CO and optical/near-IR searches for galaxies associated with high-$z$ DLAs are complementary, with the former sensitive to the more massive, dusty galaxies associated with the highest-metallicity DLAs, and the latter to the lower-mass and less dusty galaxies. Our results further indicate that the \hi-absorption selection yields the {\it complete} population of galaxies at high redshifts, $z \gtrsim 2$, and a diversity of environments, including dusty objects that are bright in CO and dust emission, and dust-poor objects that are bright in stellar emission. The \hi-absorption selection also yields the identification of high-$z$ galaxy groups, where the damped Lyman-$\alpha$ absorption may arise from neutral hydrogen lying in between the galaxies \citep[e.g.][]{chen19,peroux19}.

In summary, we have carried out an ALMA search for CO(3$-$2) or CO(4$-$3) emission in the fields of 12 high-metallicity 
([M/H]~$\geq -0.72$) DLAs at $z \approx 1.7-2.6$. We detect CO emission from five galaxies at redshifts in good agreement
with the DLA redshift; the impact parameters of the CO emitters to the QSO sightline are $\approx 5.6-100$~kpc. 
We obtain high molecular gas masses, $\mmol \approx (1.3-20.7) \times (\alpha_{\rm CO}/4.36) \times 10^{10} \msun$, 
with the highest CO line luminosities and inferred molecular gas masses arising in galaxies associated with the highest 
metallicity DLAs, [M/H]~$\gtrsim -0.3$. The good agreement between the CO emission redshift and the metal-line absorption 
redshift in all four DLAs with impact parameters $\lesssim 30$~kpc suggests that the DLAs may arise in gas clumps in the CGM 
of the galaxies, although absorbers at low impact parameters $\lesssim 10$~kpc may contain contributions from the disk of the 
CO-emitting galaxy. The fifth DLA, towards J0918+1636, is likely to arise in gas associated with a spectroscopically-confirmed 
galaxy with a lower impact parameter. The high molecular gas masses of the galaxies associated with the five DLAs can be plausibly 
explained by a combination of the stellar mass-metallicity relation and a high molecular gas-to-stars mass ratio in high-$z$ galaxies.

\acknowledgements
NK acknowledges support from the Department of Science and Technology via a Swarnajayanti Fellowship 
(DST/SJF/PSA-01/2012-13), and from the Department of Atomic Energy, under project 12-R\&D-TFR-5.02-0700. MN acknowledges support from ERC advanced grant 740246 (Cosmic{\verb|_|}Gas). 
JXP acknowledges support from NSF AST-1412981. LC is supported by 
DFF - 4090-00079. Support for this work was provided by the NSF through award SOSPA2-002 
from the NRAO. ALMA is a partnership of ESO (representing its member states), NSF (USA) and 
NINS (Japan), together with NRC (Canada), NSC and ASIAA (Taiwan), and KASI (Republic of Korea), 
in cooperation with the Republic of Chile. The Joint ALMA Observatory is operated by ESO, AUI/NRAO 
and NAOJ. The data reported in this paper are available though the ALMA archive 
(https://almascience.nrao.edu/alma-data/archive) with project codes: ADS/JAO.ALMA \#2016.1.00628.S 
and \#2017.1.01558.S.


\end{document}